\documentclass[conference]{IEEEtran}
  \usepackage{cite}
\ifCLASSINFOpdf
\else
\fi
\usepackage[pdftex]{graphicx}
\usepackage[export]{adjustbox}

\begin{document}
\title{A Token-Based MAC Solution for WiLD Point-To-Multipoint Links}

\author{\IEEEauthorblockN{Carlos Leoc\'{a}dio, Tiago Oliveira, Pedro Silva, Rui Campos, Jos\'{e} Ruela}
\IEEEauthorblockA{INESC TEC and and Faculty of Engineering, University of Porto\\
Rua Dr. Roberto Frias, 378\\
4200-465 Porto\\
Email: \{cmsl, ttpo, pmms, rcampos, jruela\}@inesctec.pt }
}


% compsoc conferences compared to traditional conferences), use this
%Telephone: (800) 555--1212, Fax: (888) 555--1212}


\maketitle

\begin{abstract}
The inefficiency of the fundamental access method of the IEEEE 802.11 standard is a well-known problem in scenarios where multiple long range and faulty links compete for the shared medium. The alternatives found in the literature are mostly focused on point-to-point and rarely on long range links. Currently, there is no solution optimized for maritime scenarios, where the point-to-multipoint links can reach several tens of kilometers, while suffering from degradation due to the harsh environment. This paper presents a novel MAC protocol that uses explicit signalling messages to control access to the medium by a central node. In this solution, this role is played by an Access Point that sends token messages addressed to each associated station, which is granted exclusive access to the medium and assigned a number of credits; after sending its own packets (if any), a station must release control of the token. Simulation results show that this mechanism has superior performance in point-to-multipoint, long range and faulty links, allowing a fairly and more efficient usage of the shared resources among all nodes.

\end{abstract}


% \fi
\IEEEpeerreviewmaketitle

\IEEEoverridecommandlockouts


\section{Introduction}

The IEEE 802.11 standard \cite{[1]} specifies a Distributed Coordination Function (DCF) as the basic medium access control (MAC) method. DCF uses a contention-based carrier sense multiple access with collision avoidance (CSMA/CA) protocol, which was designed for short-range communications scenarios. According to DCF, a node must sense an idle medium for a DIFS interval before transmitting a frame. If during this time the medium is sensed busy, the node defers its transmission and calculates a back-off interval, according to a binary exponential function, and starts transmitting only when the back-off timer runs out (this timer is paused during busy periods), even if the medium becomes idle in the meanwhile. The back-off time is a multiple of the slot time, a system parameter that is increased by 1 $\mu$s for each 300 m increase of the link length to account for the increase of the round-trip delay. 

The inefficiency of the protocol is due to three factors: 1) frame losses because of collisions (aggravated by the hidden node problem) and erroneous frames (due to channel impairments and interference phenomena in the ISM band), which trigger MAC layer retransmissions; 2) waiting time for a round-trip interval until an acknowledgement is received or a time-out occurs (depending on the protocol version, single or block ACKs are used, and thus different retransmission procedures may be invoked); 3) the back-off interval, which doubles on average for each retransmission attempt. The negative impact of these factors on the achievable throughput worsens in long-range wireless links. 

Moreover, in an access network composed of N nodes – one Access Point (AP) and N-1 client stations (STAs) – the CSMA/CA mechanism limits the AP to only acquire 1/N of the channel’s resources when all nodes contend for the medium, since the AP is treated as any station and may even be penalized by the back-off mechanism. This creates contention asymmetry between uplink and downlink traffic, which is even more undesirable since, nowadays, there is also an asymmetric traffic demand, with most of Internet traffic (e.g., Web browsing and streaming applications) being sent on the downlink direction.

In a maritime environment, the Internet access is often limited to the usage of cellular networks near-shore or the expensive alternative of satellite communications in remote ocean areas. In \cite{[2]} and \cite{[3]} the authors identify the need for a cost-effective sea communications solution to be used by the fisheries community. According to \cite{[4]}, a system capable of accommodating high-throughput communications would allow extending the concept of Internet of Things (IoT) to the sea environment (Internet of Things at Sea), to support off-shore monitoring, remote ship operations and fleet management.

Over the years, there have been several attempts to extend Wi-Fi coverage to the sea. However, in such scenarios, the Wi-Fi links can reach up to tens of kilometers and are highly sensitive to the unstable propagation conditions, which means that the use of DCF significantly degrades performance (higher packet loss ratio and delay and lower throughput), furthermore when considering point-to-multipoint links. Replacing DCF with a more efficient and fair mechanism should allow achieving significant gains in the utilization of the channel capacity (thus, on throughput) and, desirably, a more controlled sharing of resources among flows.

The main contributions of this paper are the proposal of a new medium access protocol (based on a token managed by the AP) targeted at these challenging scenarios, and its validation and evaluation by means of ns-3 simulation.

%The rest of the paper is organized as follows. Section II discusses the related work on medium access control mechanisms, including those also based on the token concept. Section III details the architecture of the proposed solution (that, nonetheless, reuses some MAC functions of IEEE 802.11). Section IV presents the simulation setup, while Section V analyses the simulation results, aiming at the performance evaluation of the token mechanism and its comparison with DCF. Section VI gives a brief outline of the ongoing development of a laboratory prototype and a follow-up pilot demonstrator. Finally, Section VII draws the main conclusions and points out the planned future work.

\section{Related Work}
\label{related}

%The IEEE 802.11[1] standard has evolved over time, with the inclusion of many extensions and amendments. In particular, technological advances in modulation and coding schemes as well as new antenna designs have allowed an impressive increase in achievable data rates. 

%The efficiency of IEEE 802.11 depends on the relative weight of PHY/MAC overhead, due to PHY framing and MAC contention, when compared with the time to transmit useful data. In IEEE 802.11g both overhead components are accounted for on each frame. Thus, the efficiency decreases with short frames as well as when extending the network range or increasing the data rates. These problems are attenuated in IEEE 802.11n with both frame aggregation and BACKs, since the PHY/MAC overhead is shared by multiple frames in a single access (with a unique round-trip time per BACK).  

%However, with the continuous increase of data rates, the amount of aggregated data should increase proportionally to keep the same benefit. While this may be acceptable for bulk delay tolerant data, it is not feasible for real-time traffic (due to the higher delays to collect data) or applications that generate small amounts of bursty data or short frames.

The IEEE 802.11 inefficiency problem has been addressed by new schemes that adopt the standard (with no or minor modifications) but with a different organization of physical channels. Instead of sharing a single wide channel among nodes, an alternative is to split spectrum into multiple narrow channels to increase spectral efficiency. Decreasing the data rate on each channel extends the time to transmit a given amount of data, thus reducing the relative PHY/MAC overhead and achieving the same goal as aggregation with much shorter frames. 

Fine-Grained Channel Access (FICA) \cite{[5]},  WiFi-NC \cite{[6]} and Fine-grained Spectrum Adaptation (FSA) \cite{[7]} are three examples of fine-grained, multi-channel schemes, where the IEEE 802.11 medium access protocol is used on each channel, which is shared by all nodes. Despite spectral efficiency gains, they all suffer from the DCF limitations. Moreover, they will not likely be implemented using commercial off-the-shelf cards since they require dedicated, complex and more expensive hardware. 

Virtual Duplex \cite{[8]} addresses the contention and traffic asymmetry problems, taking a different approach. The available spectrum is divided in two independent bidirectional channels, with each channel only carrying data frames in one direction: download (AP to STA) and upload (STA to AP). IEEE 802.11 is used in both channels; although the download channel is collision-free in a single AP scenario, contention still exists in the upload channel, which is shared by the client stations. The traffic asymmetry problem is solved by statically configuring channel bandwidths to match the expected traffic loads offered to both channels, thus making the solution unable to adapt to traffic changes that occur on short time scales. 

These FDD schemes do not address other IEEE 802.11 problems, such as hidden and exposed nodes, exponential back-off and the capture effect. A different line of approach proposes replacing the 802.11 MAC with a TDMA mechanism, in order to improve channel efficiency and achieve fairness. 

The 2P MAC protocol \cite{[9]} was developed for long-distance mesh networks; it reuses 802.11 hardware but disables the generation of ACKs and the carrier sense back-off. Each node has multiple radios (one per link) that share the same channel on the whole network using directional antennas and a synchronous operation of its point-to-point links, which alternate between transmit and receive slots of fixed length. When a node switches from transmitting to receiving, its neighbours must switch the opposite way and vice-versa.

WiLDNet \cite{[10]} builds upon 2P, proposing an architecture targeting wireless long distance (WiLD) point-to-point links, using a contention-free TDMA-based medium access, an implicit loose time synchronization approach and an adaptive loss-recovery mechanism that combines forward error correction with bulk acknowledgments and retransmissions. 

JazzyMAC \cite{[11]} addresses limitations of 2P and WiLDNet architectures. It introduces the concept of variable-length time slot, whose size is dynamically chosen by each node based on a set of local parameters. The mechanism is fully distributed and allows the system to adapt to traffic demand variations. Loose synchronization is achieved by the exchange of token messages. 

The work presented in \cite{[12]} describes a centrally controlled solution targeted at generic multi-hop mesh networks, using a frame formed with three types of variable size time slots for control, contention and data. This solution requires a tight multi-hop time synchronization.

In \cite{[13]} the authors propose WiLDToken, a token-based MAC protocol that borrows concepts from 2P and Jazzy and is aimed at increasing throughput, reducing delay and achieving fairness. The paper is focused on the token exchange in a single long-distance link, which is part of a multi-radio multi-channel mesh network. The protocol state machine is divided in two phases, respectively for synchronization and data transmission. 

Unlike the previous solutions, JaldiMAC \cite{[14]} aims at optimizing long distance wireless point-to-multipoint links. It handles traffic asymmetry adaptation and provides loose quality of service guarantees for latency sensitive traffic as well as the fair usage of the spectrum resources. It uses centralized polling to assign time slots to stations; at the beginning of each round, stations use a contention slot to request access to the medium. To avoid collisions, the contention slot must increase with the number of stations, which may reduce the overall efficiency.

The solution described in this paper addresses and tries to overcome most of the above mentioned problems. On one hand, it reuses off-the-shelf IEEE 802.11 hardware, only retaining the PHY framing structure as well MAC aggregation and BACK mechanisms. On the other hand, DCF contention (including exponential back-off) is eliminated and replaced with a centralized token scheme that takes advantage of the underlying functions. Although for validation and performance evaluation only a small set of basic features has been selected, the next section presents a general framework of the token mechanism and how its flexibility and advanced features may be exploited and tuned to meet different performance and fairness objectives.

\section{Architecture}
\label{architecture}
The proposed protocol is based on the token concept. The circulation of a single token among nodes is centrally controlled by a Token Manager that runs on the AP, which assigns transmission opportunities to each station (STA) by means of an explicit signalling message that contains the token. The AP also schedules its transmissions when owning the token and is responsible for maintaining an updated list of associated STAs.

Collisions are avoided by ensuring that, at every instant, only the node holding the token is allowed to transmit data frames. Efficiency is improved at high loads by eliminating the overhead due to collisions and retransmissions (and associated delays). The protocol procedures require Data and Control channels.

In the Data channel, the token protocol replaces the CSMA/CA based access method. Besides signalling messages exchanged between the AP and STAs, the protocol must include recovery mechanisms to prevent deadlocks in case of loss of some messages. However, the IEEE 802.11 acknowledgment mechanism of data frames is kept as well as the PHY and MAC frame structure. The protocol works with either individual data frames and ACKs or aggregated frames and BACKs. In this paper, frame aggregation is exploited and therefore we assume hereafter that MAC data frames are acknowledged in blocks.

The Control Channel works on a standard IEEE 802.11 link, to allow new stations to associate to the AP; moreover, this channel may also be used to broadcast emergency or other relevant information. A station is initially tuned to the Control channel and switches to the Data channel as soon as the association procedure completes with success; it may switch back to the Control channel to handle exception conditions.
The rationale and a general framework of the token protocol is discussed next, followed by details of the procedures in the Data and Control channels, frame aggregation and the token state machine.

\subsection{The token protocol – rationale and framework}
The token protocol was designed for a wireless access network with N nodes (a single AP, connected to an infrastructure network, and N-1 STAs), and a scenario where all traffic is exchanged between the AP and each STA.  

While an STA sends all its traffic to the AP, modelled as a single upstream flow, the AP has to send its traffic to all STAs, as N-1 independent downstream flows. In general, rates per unidirectional flow are variable, with different average and peak values and, for a given (AP, STA) pair, traffic is asymmetric (typically in favour of the downstream direction, considering current applications). For these reasons, the amount of traffic that must be transmitted by the AP is usually much higher than the amount of traffic offered per STA.

It is, thus, reasonable to assign access opportunities per flow rather than per node, unlike DCF that, in theory, treats nodes in equal terms, irrespective of the amount of traffic they have to send, and thus penalises the AP, under high loads, in the considered traffic scenario. Moreover, a centralised token protocol combined with a credit mechanism managed by the AP allows controlling individual flow rates taking into account the offered traffic per flow and some fairness criterion.

The simplest way is to organise transmissions in rounds during which the AP selects the STAs in a round-robin manner to assign them the token with a credit value. Each time an STA receives the token, it is granted exclusive use of the medium to transmit packets (within its quota), after which it must return the token to the AP, at the earliest opportunity. 

Since the AP regains control of the token each time it is released by a station, it can decide when and where to transmit its packets. In order to exploit the aggregation of frames per downstream flow and allow a tight control of the respective rate, the AP must organize frames per flow (or destination STA), and not by order of packet arrival. Serving each flow once in a cycle gives upstream and downstream flows one opportunity per round, while controlling the flow rates by means of the amount of individual credits assigned per round. Although it would be possible to decouple the upstream from the downstream transmissions, a simple way of accomplishing the above idea would be for the AP to select a station, transmit packets to that station, followed by the token, and then prepare to receive packets and the token back from that station.

Besides controlling the individual flow rates in a fair manner, it is also necessary to control the round duration, which has impact on the delay experienced by the flows. A system goal might be setting a maximum target round time based on a performance objective.

The round duration has a fixed component (the overhead due to the transmission and processing of the token and the total round trip delay due to the circulation of the token) and a variable component (the time to transmit data frames and BACKs and the associated round trip delay per BACK). This variable component contributes to the overall delay, and will depend on the criteria to assign credits to flows. A basic criterion would be, for example, to assign a maximum number of packets that a flow can transmit per access.

This strategy could be combined with another mechanism, by extending the round concept such that, in a round, some flows might be given more than one opportunity, while still controlling the maximum round time. This may be interesting since it is a means of assigning additional credits per round to some flows, without exceeding the maximum number of credits per access. In this way, it is still guaranteed that all flows have at least one transmission opportunity and thus a guaranteed access delay.

As outlined, the flexibility of the token mechanism allows extending the protocol with advanced features, with different degrees of complexity. It is possible to tune credit assignment to the “instantaneous” flow rates (averaged over some past temporal window) and treat saturated flows according to a configurable weighted max-min fairness criterion.

Although we have set a broad framework for the token mechanism, it is beyond the scope of this paper to deal with all these aspects. The main goal is to compare the basic token protocol with DCF, identify its main advantages and possible weaknesses, and derive some useful conclusions about critical parameters and how to tune them.

\subsection{Control Channel}
The Data channel carries Data frames and BACKs (using IEEE 802.11 standard procedures) as well as signalling messages of the token protocol. The credit assigned to each STA is included in the token message and, in a dynamic scheme, may vary between a default and a maximum value, as explained. 

Upon receiving the token, an STA verifies whether it has data to transmit and, if so, it starts transmitting until all credits are consumed or there is no more data to send, whatever happens first. Then, it returns the token to the AP, optionally reporting the amount of credit used, which allows the AP to create a profile, in real time, of the traffic needs of each STA, to help the assignment of future credits. When an STA has no data to transmit, it immediately releases the token.

Similarly, after regaining control of the token, the AP initiates transmission of frames of a particular flow (addressed to a specific STA), up to the assigned credit. 

The token is not released (either by the AP or an STA) before an expected BACK to outstanding frames is received or an associated timer expires. Moreover, the existence of a live and valid token is also protected by a timer in the AP to trigger system recovery in case it is not returned by an STA, due to transmission errors or failure of the STA.

The sequence of operations of the Token Manager is illustrated in Figure 1.

\begin{figure}
	\centering
	\includegraphics[scale=0.54]{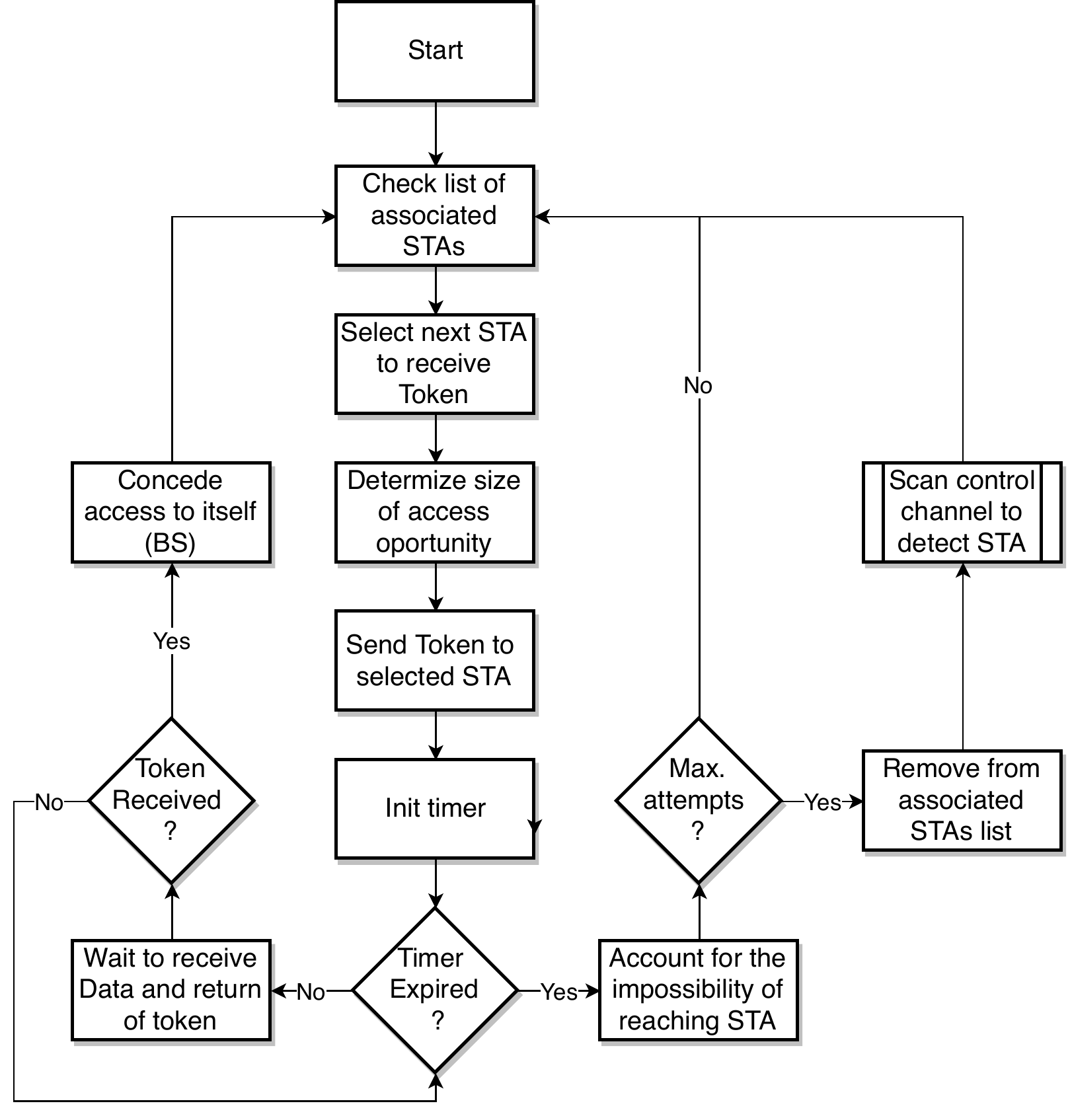}
	\caption{Token manager sequence operation}
	\label{digrama}
\end{figure}

Performing the association procedure through the token managed Data channel proves to be impractical, since an STA is not initially associated with the AP and therefore it will never be given the opportunity to access the medium to start the association process.

To overcome this problem it is necessary to devise a mechanism that allows the association of new STAs to the AP, or even their re-association, since the probability of losing a connection is not negligible in the harsh maritime environment. The proposed solution makes use of a Control channel, based on the IEEE 802.11 standard and, thus, operating according to the DCF mechanism, which is acceptable since the load on this channel is low.  

Any STA that wishes to join the access network, should first associate with the AP through this channel. The STA must identify itself and request access to the Data channel. If the association procedure succeeds, the STA is notified that it must switch to the Data channel and, from this moment on, the table of clients associated with the AP includes the new STA, which will be eligible to receive the token according to the AP schedule.

If an STA detects that it may have lost its association with the AP, because it has not received the token within a maximum specified interval, it must switch back to the Control channel, and initiate a new association procedure. In the meanwhile, the traffic generated by the STA should be discarded, until a successful re-association completes.

The existence of a Control channel requires the AP to have two radio interfaces, permanently active, one for the Control channel and another for the Data channel. On the other hand, to reduce costs, the STAs only require one radio interface, which will switch between both channels, as explained. In the proposed solution, the Control channel should be configured with a bandwidth of 5/10 MHz to provide greater range and better noise immunity, at the expense of throughput. The longer range of the Control channel makes it particularly suitable for broadcasting information of common interest (e.g. meteorology) or for emergency relief, and is therefore the default channel of operation.

\subsection{Frame Aggregation}
\label{fa}
The proposed architecture transparently uses the existing aggregation mechanism specified in the IEEE 802.11n standard.

Considering the communication between the AP and a given STA, it is necessary to independently negotiate frame aggregation for transmissions in both directions. A handshake procedure, based on the exchange of management frames, must be performed for each direction: the sending node issues an ADDBA Request and, if the receiving node is willing to accept aggregated frames, it answers with a positive ADDBA Response, and both nodes enable the aggregation mechanism for transmission and reception, respectively.

Since this is management procedure of the request-reply type, it must be performed independently and not interfering with the normal transmission of data frames, which is controlled by the token, with a different semantic. The node that initiates the handshake must be in possession of the token (to avoid collisions) but should not release it to the receiving station, which is implicitly allowed to access the medium to immediately respond to the sender request. This procedure must be protected by a timer and possibly be repeated a number of times (immediately or later), if a response is not received. In case the negotiation is unsuccessful, frame aggregation is not possible in the corresponding direction. 

At the end, the sender, as owner of the token, decides on the next action, according to the normal data transfer procedures.

\subsection{Token State Machine}
The token operation is described by the state machine represented in Figure 2, which shows the possible events and conditions that trigger state transitions.

\begin{figure}
	\centering
	\includegraphics[scale=0.60]{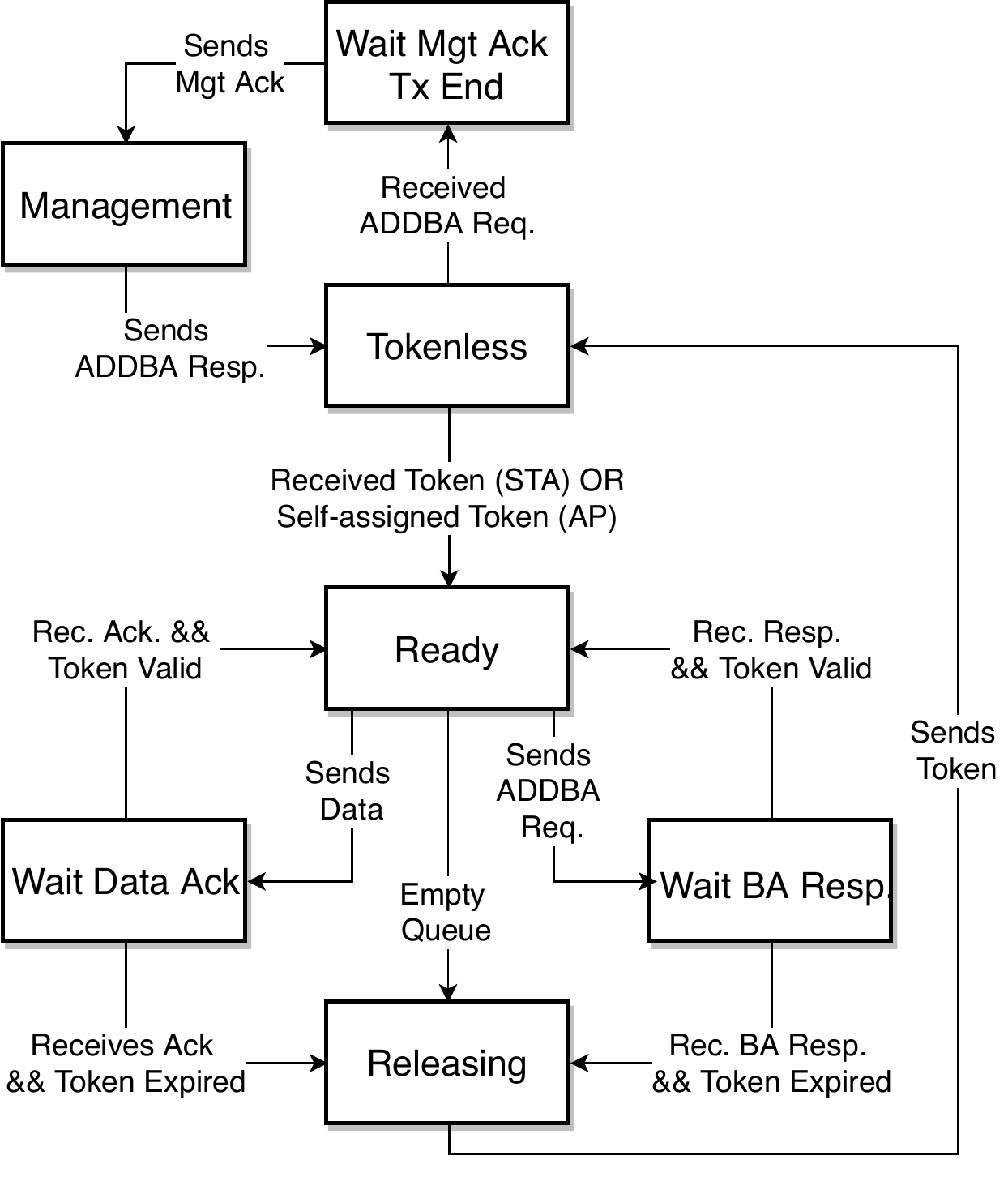}
	\caption{Token state machine}
	\label{stateM}
\end{figure}

When a node does not own the token, it is in the Tokenless state. Once it receives the Token (STA) or self-assigns it (AP), a transition to the Ready state takes place, which means the node is allowed to access the medium. When the node finishes transmitting an aggregated frame (A-MPDU), it will wait for a layer 2 BACK in the Wait Data Ack state and, upon reception, will move back to the Ready state, if there are still credits available. When the node runs out of credits or has no more packets to send, it will enter the Releasing state, to return the token to the AP (in case it is an STA) or to pass it to another node (if it is the AP). After sending the token, the node is back in the Tokenless state. 

In the Ready state a node may also initiate the negotiation of the aggregation mechanism, while in the Tokenless state it may be invited to accept aggregated frames. Although the transmission of ADDBA management frames is protected by layer 2 ACKs, the loss of these ACKs is irrelevant (and should not trigger a retransmission), since the success of the negotiation from the point of view of the requester is confirmed by the reception of an ADDBA Response to an ADDBA Request. 

When a node receives an ADDBA Request, it will move to the Management state in order to send the ADDBA Response to the originator, after being delayed until transmission of the layer 2 ACK completes, and will return to the Tokenless state once the response is sent. As an originator, the node sends an ADDBA Request in the Ready state and waits for a response in the Wait BA Resp state, only switching back to the Ready state when the response is received; the node holds the token during this procedure.

\section{Simulation Setup}
\label{sim}

The token protocol was implemented on the discrete-event network simulator ns-3, in order to assess and compare its performance with the standard DCF access method. The ns-3 Wi-Fi module can be roughly split in 3 submodules: PHY layer, MAC low and MAC high. The DCF medium access mechanism is modelled in the MAC low module by means of the DcfManager model. For the token protocol, this model was deactivated and replaced with the TokenManager model. For evaluation and comparison, only the Data channel procedures are required and thus it was assumed that stations were already associated with the AP. 

The simulation setup is represented in Figure 3, with 10 nodes (N = 10) with the stations at increasing distances from the AP, such that for STA$_i$ the distance is i*D, with D = 500 m. The nodes were equipped with 2.4 GHz 802.11n based radios, on a bandwidth of 20 MHz, using 20 dBm of transmission power and an antenna gain of 10 dBi.

\begin{figure}
	\centering
	\includegraphics[scale=0.45]{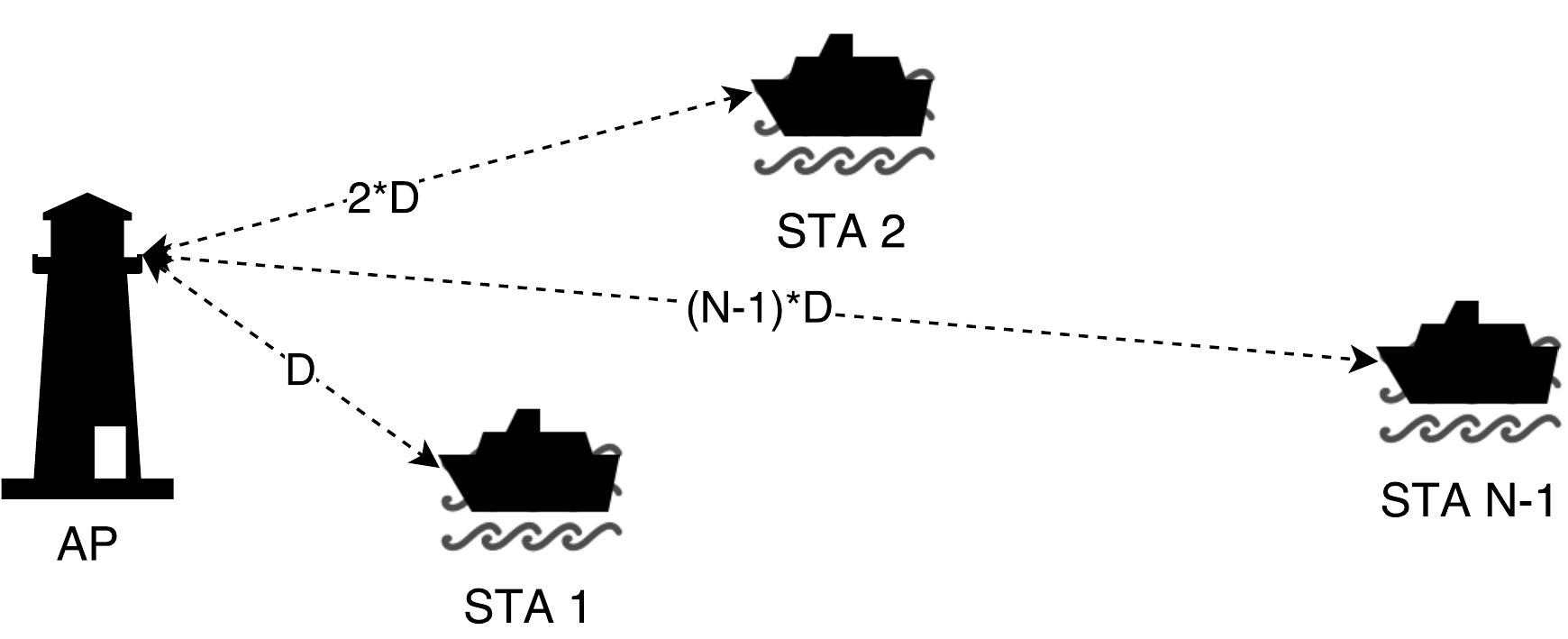}
	\caption{Simulation scenario}
	\label{scenario}
\end{figure}

Two UDP constant bit rate (CBR) flows were established between the AP and each STA (one per direction). Two generation rates of 1000 or 2000 kbit/s were considered, in order to evaluate the performance of both access methods; the lowest rate was chosen such that the network is operating below saturation in the case of the token protocol. The impact of using an adaptive rate manager (Minstrel HT) compared with the use of a fixed channel rate was also assessed; in the latter case, the MSC4 modulation and coding scheme (with a 39 Mbit/s channel rate) was chosen for all cases, to account for the highest distance to the AP. Packets are 512 bytes long. 

The IEEE 802.11n standard was selected to take advantage of the frame aggregation mechanism and thus improve performance. The maximum number of MPDUs (MAC frames) within an A-MPDU (aggregated frame) was limited to 4, instead of the default 64 frames, in order to prevent a node from occupying the medium for a too long time, while sending a big A-MPDU. 

The maritime environment is prone to cause fluctuations on the signal level, due to the constant changes on the antenna height and direction (because of ship oscillation), as well as the different reflections that the signal suffers on the waves, which can be either constructive or destructive to the main ray. In order to model such variability, a Rician fading component was used on top of the standard Friis propagation model.

A total of four scenarios were simulated, combining the two generation rates and the two rate modes (constant and adaptive), as shown in Table 1.

\begin{table}[htbp]
	\caption{Scenarios evaluated}
	\begin{center}
		\begin{tabular}{|c|c|c|c|c|}
			\hline
			\textbf{Scenario} & \textbf{N} & \textbf{D (m)} & \textbf{Data Rate (kbps)}& \textbf{Rate Manager} \\
			%\cline{2-4} 
			%\textbf{Head} & \textbf{\textit{Table column subhead}}& %\textbf{\textit{Subhead}}& \textbf{\textit{Subhead}} \\
			\hline
			1 & 10 & 500 & 1000 & Constant (MCS4) \\
			\hline
			2 & 10 & 500 & 1000 & Minstrel (HT) \\	
			\hline
			3 & 10 & 500 & 2000 & Constant (MCS4) \\
			\hline
			4 & 10 & 500 & 2000 & Minstrel (HT) \\
			\hline
		\end{tabular}
		\label{tab1}
	\end{center}
\end{table}

The simulations run for 500 s with 30 different seeds. The flows are randomly initiated within an interval of 2 s, differently for each seed.

\section{Simulation Results}

In the following analysis, average values per flow considering all simulation runs with different seeds are given. However, to illustrate particular behaviours and differences between upstream and upstream flows, values obtained with seed 10 are used.

A summary of the main results that will be used in the following analysis is given in Table 2.

The token and DCF schemes are separately analysed, first considering a fixed channel rate (scenarios 1 and 3). Then a brief analysis of the adaptive scheme is provided (scenarios 2 and 4). Throughout this section, when relevant, performance differences among all cases will be highlighted.

\begin{table}[htbp]
	\caption{Average Flow IP Statistics}
	\begin{center}
		\begin{tabular}{| c | c | c | c | c | }
			\hline
		%	&multicolumn{4}{|c|}{\textbf{Throughput (kbps)}}  \\
		%	\cline{2-5} 
			\textbf{Scenario}&\multicolumn{2}{|c|}{\textit{Downstream}} 
			&\multicolumn{2}{|c|}{\textit{Upstream}} \\
			\cline{2-5} 
			& \textbf{DCF}& \textbf{Token} & \textbf{DCF}& \textbf{Token} \\
			\hline
			1 	
			&  587.22 &	999.90 & 716.63 & 1000.00
			\\
			\hline
			2 
			&  570.18 &	958.18 & 572.49 & 999.73
			\\	
			\hline
			3 
			&  344.10 &	1066.47 & 1004.33 & 1151.75
			\\
			\hline
			4 
			&  482.71 & 1026.31 & 780.12 & 1192.25
			\\
			\hline			
		\end{tabular}
		\label{tab2}
	\end{center}
\end{table}

\subsection{Token}

\subsubsection{Scenario 1}

The offered traffic does not saturate the network, since it is possible to successfully transmit at the MAC layer all generated UDP packets, except for some negligible losses in the downstream direction, in spite of the need to retransmit frames. Those losses (in the order of 0.01\%) are due to dropping frames that exceed the 500 ms time to live (TTL) in the MAC queue. The throughput of each upstream flow is equal to the generation rate (1 Mbit/s), while in the downstream direction the average is 999.9 kbit/s.

It is worthwhile mentioning some differences observed in the behaviour of the flows, which are not relevant in this case but may have impact in the other cases.

The number of frames that must be retransmitted increases with the flow index, that is, with the distance between each STA and the AP, thus affecting the aggregation efficiency. As an example (with seed 10), in the upstream direction, for flow 1 (with almost no retransmissions) on average 2.8 frames are transmitted per A-MPDU, while for flow 9 (with 30\% of retransmissions), this value raises to  3.7; moreover, the number of non- aggregated frames is negligible for all flows. In the downstream direction, about 4\% of frames are not aggregated; since packets of the same flow are scattered in the shared IP buffers and handled on a single queue, it happens that, at the MAC layer, some frames are not eligible for aggregation with other frames of the same flow. This problem does not occur on the upstream direction, since each STA only has to handle a single flow. On average, less than 10\% of frames have to be retransmitted in each direction.  

It was observed, for all seeds, that in the upstream direction the number of successfully transmitted frames was equal to the number of packets transferred to the MAC layer (no MAC losses), which was also equal to the number of generated packets (no IP losses). However, in some cases, the number of received packets at the IP layer was slightly different, possibly due to processing delays at the receiver beyond the end of the simulation interval. Therefore, to keep consistency with the behaviour observed at the MAC layer and ease the calculations, throughput was calculated in all scenarios considering the ratio of successfully transmitted MAC frames to the number of generated UDP packets multiplied by the respective generation rate. 

This scenario constitutes a useful reference, because it provides a simple means of calculating throughputs and allows identifying possible causes of performance degradation or unfairness in the other cases.  

\subsubsection{Scenario 3}
In this scenario, the network is saturated and thus there will be significant losses since the offered load doubled when compared with scenario 1, where the network was already on de edge saturation, thus with little margin for improvement. It is now possible to assess how fairly flows are treated, in what concerns overall throughputs in both directions and individual throughputs in each direction.

The loss behaviour of the downstream flows is difficult to predict, since it depends on the interaction with the MAC layer. As soon as MAC buffers are fully occupied, transfer of packets from the IP to the MAC layer will only be possible when successfully transmitting MAC frames and thus freeing the corresponding buffers (losses at the MAC layer will occur only when frames are not successfully (re)transmitted or their TTL is exceeded). Thus, the IP queue will build up until all IP buffers are filled and then new packets will only be accepted when transferring older packets to the MAC layer. However, since frame retransmissions occur at random and vary with link lengths, freeing MAC buffers will occur at uneven (irregular) intervals. It is expected that most losses will occur at the IP layer and acceptance of new IP packets will not be correlated with the flow indices, thus creating random differences in the individual throughputs.

In the upstream direction, as expected, the total number of transmitted frames per flow (including retransmissions) is approximately the same, since flows are generated on different stations and thus equally treated by the protocol. Since the number of retransmissions increases with the flow index, the respective throughputs decrease. In the downstream direction, the throughput per flow does not exhibit this predictable pattern; in fact, simulation results confirm that individual throughputs are not correlated with flow indices and that flow patterns vary with different seeds. However, the total downstream throughput is approximately the same with different seeds, due to the regulatory effect of the protocol.

In the upstream direction, the number of non-aggregated frames is negligible and the aggregation efficiency is high, close to 100\% (near four frames per A-MPDU). In the downstream direction, the number of non-aggregated frames was in the order of 1\% (an improvement over scenario 1) and the aggregation efficiency is also high (only slightly lower than in the reverse direction). Global losses (measured at the IP layer) are in the order of 47\% in the downstream direction and 42\% in the upstream direction, with corresponding average throughputs of 1055 and 1150 kbit/s. The losses at the MAC layer in the downstream direction are in the order of 0.01\% and much lower in the upstream direction, thus confirming the analysis.

The behaviour in the downstream direction is not due to sharing the buffer space among nine flows, each one getting on average one ninth of the buffers available for upstream flows, but because packets are handled on a single queue. In a real implementation, this could be solved using separate queues per flow in the shared buffer space. However, this question deserves further discussion. In a lasting saturation condition (as in scenario 3), increasing the number of buffers does not reduce the packet loss ratio and only increases the average end-to-end packet delay. The figures for packet delays can thus be misleading if not crossed with packet losses and number of buffers. For this reason, they are not presented in Table II, but are given here to justify this reasoning. 

The average packet delay in the upstream direction is in the order of 3.9 s; this is consistent with the time that packets at the tail of a permanently fully queue of 1000 buffers have to wait until reaching the head of a queue emptied at the useful rate of 1150 kbit/s. In the downstream direction the same 1000 buffers are shared by nine flows; thus, the single queue is emptied at a rate that is nine times the average useful rate of 1055 kbit/s, with a lower average packet delay of around 470 ms. In conclusion, increasing the number of buffers is not a solution to reduce packet losses (and thus increase the throughput), and is only useful to solve transient peak loads, at the expense of increasing the delay during such periods. This issue will be reanalysed in the DCF cases, where it appears with a different flavour.

Since in this scenario the network is saturated (with an aggregation efficiency close to 100\% considering all transmission attempts), we may conclude that, under the imposed conditions, the useful network capacity has been reached. This means that, considering all nine bidirectional flows, it is possible to achieve an average throughput of 19.845 Mbit/s, that is, 50.9\% of the physical channel rate. The following components contribute to the total overhead (49.1\%) is due to the following causes: 1) PHY framing (that increases with non-aggregated frames); 2) MAC acknowledgment mechanism (BACK/ACK framing and round-trip time); 3) frame retransmissions; and 4) circulation of the token (token frames and ACKs and associated round-trip time). Only the last component is accountable to the token mechanism. This suggests some possible ways to increase the efficiency:

\begin{itemize}
	
\item increase the number of credits per access, which increases the average cycle time and thus the delay. 
\item better tune the physical rates of the different links, either statically or dynamically (avoiding too frequent changes), which requires evaluating the trade-off between more retransmissions at higher rates and more time for frame transmissions at lower rates; 
\item embed the token signal in the header of the last frame of an access, which may create other problems that should be carefully evaluated due to the interaction with the normal acknowledgment process (for example, need to retransmit frames other than the frame that contains the token). 

\end{itemize}

\subsection{DCF}
\subsubsection{Scenario 1}
Unlike the token scheme in scenario 1, the network is now congested, because collisions contribute to a higher number of retransmissions that add to those due to transmission errors. This means that queues will build up at the IP layer and packets will be lost.

Simulation results show that on the downstream direction the average loss ratio is around 40.3\% while it is 28.4\% on the upstream direction; the average throughputs are 587.2 and 716.6 kbit/s, respectively.  These results show a poorer performance of DCF in comparison with the token method, but hide the unfair treatment of individual flows. Looking into detail at the flow level, we gain another insight into the problems, which are illustrated with a particular instance of simulation (using seed 10).

In the downstream direction, except for the higher loss ratio and thus a lower throughput, it was observed, as in the token cases, that there is no correlation between individual throughput and flow index and the differences in throughputs are not severe.

In the upstream direction, the unfairness is noticeable, with a capture effect by some flows. The throughput of six flows almost reached the generation rate (1 Mbit/s) while the remaining three flows were virtually starved; such starvation does not occur for the flows with higher indexes, which get almost all their packets transmitted in spite of requiring a higher number of retransmissions. With other seeds, the individual behaviour is different, but the global behaviour is similar.

The aggregation figures are rather different for both directions. The downstream flows benefit from an aggregation efficiency close to 100\% (considering frame retransmissions) and a small number of non-aggregated frames (around 0.2\% of all successful frames). In the upstream direction, not only the aggregation efficiency is lower but also the number of non-aggregated frames is much higher. As an example, for seed 10, considering increasing flow indices and excluding the starved flows, the average number of aggregated frames per A-MPDU ranges from 2.2 to 3.0 and the percentage of non-aggregated frames ranges from 91.5\% to 41.3\%. This explains why contention asymmetry in favour of the upstream direction is moderate.

The average delay measured at the IP layer in the downstream direction is around 890 ms, which is consistent with the buffer size and average throughput. In the upstream direction the average delay is low (in the order of 10 ms), but this value is quite misleading, since packets that are successfully transmitted have short delays while starved flows do not get almost all their packets transmitted. The latter even contribute to reducing the average delay (over all nine flows), since their weight in the averaging process is almost zero. This is rather different from the token case and explains the huge difference observed. 

Some of these problems become even more evident and severe in scenario 3. 

\subsubsection{Scenario 3}
In this case, the packet loss ratio has a significant increase, with average values of 82.8\% and 42.4\% in the downstream and upstream directions, respectively, to which correspond average throughputs of 344.1 and 1004.3 kbit/s. The contention asymmetry is more evident than in scenario 1, with a throughput increase in the upstream direction and a decrease in the downstream direction, with only a minor increase of the total in both directions. 
 
Considering again seed 10 as an example, the throughput of five upstream flows almost reached the generation rate of 2 Mbit/s, one flow got around 1 Mbit/s and the other three were starved. In what concerns downstream flows, the same behaviour as in scenario 1 was observed, except for a major throughput reduction. The main reasons for a higher relative gain of the upstream direction were a higher aggregation efficiency (ranging from 2.7 up to 4 aggregated frames per A-MPDU) and a much lower percentage of non-aggregated frames (ranging from 46.7\% to 3.6\%). In the downstream direction, the aggregation efficiency was still close to 100\% and the percentage of non-aggregated frames raised to 0.6\% (over a much lower number of successful frames than in scenario 1). 

Comparing with the token scheme we can observe that in the token case: a) there is almost no contention asymmetry, which could be eliminated by assembling packets of each flow on separate queues; and b) flows are fairly treated (considering retransmissions), which means that flows with higher indices have a slight throughput degradation since they have to retransmit a higher number of frames.

	\label{mcs_node7}

\section{Conclusion}
\label{conclusion}
The standard medium access control mechanism (DCF) is inefficient in scenarios of long point-to-multipoint links, specially when their subject to some signal variation. In this paper, we presented a token based alternative, that uses a centralized controller to regulate the medium access between the nodes, by distributing between them tokens, which grant opportunities to access the medium.

We implemented and validated our solution using ns-3. We evaluated its performance, comparing it with the standard DCF mechanism. The obtained results proved that our solution has a better performance in this kind of scenarios, being able to reduce the packet loss, while being fairer to all nodes, even if it implied in some cases an increase of the delay.

As future work, we will carry on the on-going implementation of this solution over the Linux kernel, and then experimentally evaluate this mechanism in a long range maritime scenario.


%\ifCLASSOPTIONcompsoc
%\else
\section*{Acknowledgment}
%\fi
This work is financed by the ERDF – European Regional Development Fund through the Operational Programme for Competitiveness and Internationalisation - COMPETE 2020 under the PORTUGAL 2020 Partnership Agreement, and through the Portuguese National Innovation Agency (ANI) as a part of project MareCom (POCI-01-0247-FEDER-003468).




%\bibliographystyle{IEEEtran}

\end{document}